\documentclass[11pt,twoside]{article}

\usepackage{asp2004}
\usepackage{epsf}
\usepackage{psfig}
\usepackage{epsfig}
\usepackage{lscape}

\begin{document}
\title{Looking for Mister Goodstar (A preliminary analysis of stars with 
S~{\sc iv}, {\sc v} and {\sc vi} wind lines)}
\author{D. Massa}
\affil{SGT, Inc., Code 665, GSFC, Greenbelt, MD 20771}

\begin{abstract} 
The winds of stars with very specific temperatures and luminosities are 
ideal for determining the magnitude and nature of mass loss in OB stars.  
I identify these stars and analyze their wind lines.  The results are 
discussed within the context of recent findings which appear to indicate 
that the mass-loss rates of OB stars may as much as an order of magnitude 
less than commonly accepted values.
\end{abstract}


\section{Background}

Recent measurements of O stars winds indicate that their mass loss rates, 
$\dot{M}$, may be significantly less than previous estimates and 
expectations (e.g., Massa et al.\ 2003, Repolust et al.\ 2004, Bouret et 
al.\ 2005, Fullerton et al.\ 2006).  The primary mass loss diagnostic used 
in these analyses is P~{\sc v}~$\lambda \lambda 1118, 1128$.  Although 
wind models predict that the ion fraction of P$^{+4}$ should approach 
100\% in the winds of mid- to late O stars, this is a Na-like ion, and 
can be quite fragile.  Consequently, it is important to verify the results
based on P~{\sc v}.

\section{Sulfur Ions}

Like Phosphorus, Sulfur is a non-CNO element with a relatively low cosmic 
abundance.  Three Sulfur mass loss diagnostics are  available: the 
S~{\sc iv}~$\lambda \lambda$~1062, 1072 and S~{\sc vi}~$\lambda 
\lambda$~933, 944 resonance doublets, and the S~{\sc v}~$\lambda$~1502 
excited state line -- populated by the S~{\sc v}~$\lambda$~786 resonance 
line.  For wind conditions, the lower level of 1502\AA\ is populated 
exclusively by radiative excitation.  In this case, its radial optical 
depth in the wind is (see, Lamers et al., 1987): 
$$
\tau(v)_{rad}^1 = \frac{1}{4\pi m_H} \frac{\pi e^2}{mc} 
         \lambda_1 f_1 \left(r^2 v \frac{dv}{dr}\right)^{-1} 
	 \frac{\omega_1}{\omega_0} \frac{\lambda_0^3}{2hc} 
         \frac{\beta(v)_c}{\beta(v)} 
         \frac{A_E}{\mu} \dot{M}q(v) F[\nu_0(1 -v/c)]_{\nu} 
$$
Thus, while Sulfur provides access to three adjoining stages of ionization, 
it introduces a model dependency through the flux term, 
$F[\nu_0(1 -v/c)]_{\nu}$.

\section{Results}

S~{\sc iv}, S~{\sc v} and S~{\sc vi} occur together in luminous O4 -- O 6 
stars with massive winds.  S~{\sc v} was analyzed in both LMC and Galactic 
stars, using the SEI code (Lamers et al., 1987) and rotationally broadened 
TLUSTY models were used for both the UV and EUV continua.  The results are 
shown in Figure~1.  The derived $\dot{M}q($S~{\sc v}) should be accurate 
to better than $\pm 50$\%.

\section{Conclusions}

Preliminary Sulfur results give {\em total} sulfur mass loss rates,  
$\dot{M}q(S^{+3})+\dot{M}q(S^{+4})+\dot{M}q(S^{+5})$, between 0.04 -- 0.18 
times smaller than those expected from either theoretical (Vink et al.\ 
2000) or radio or H$_\alpha$ determinations (see, Fullerton et al.\ 2006).  
These results are similar those determined by the P~{\sc v} analyses.  
Clumping and porosity can influence the results, and must be incorporated 
into the analysis to determine exact factors.  Nevertheless, it seems that 
the mass loss rates of O stars will have to be revised downward by some 
amount. We intend to extend the current analysis to a much larger sample 
and to investigate the effects of clumping more closely.  

\begin{center}
\centering\epsfig{figure=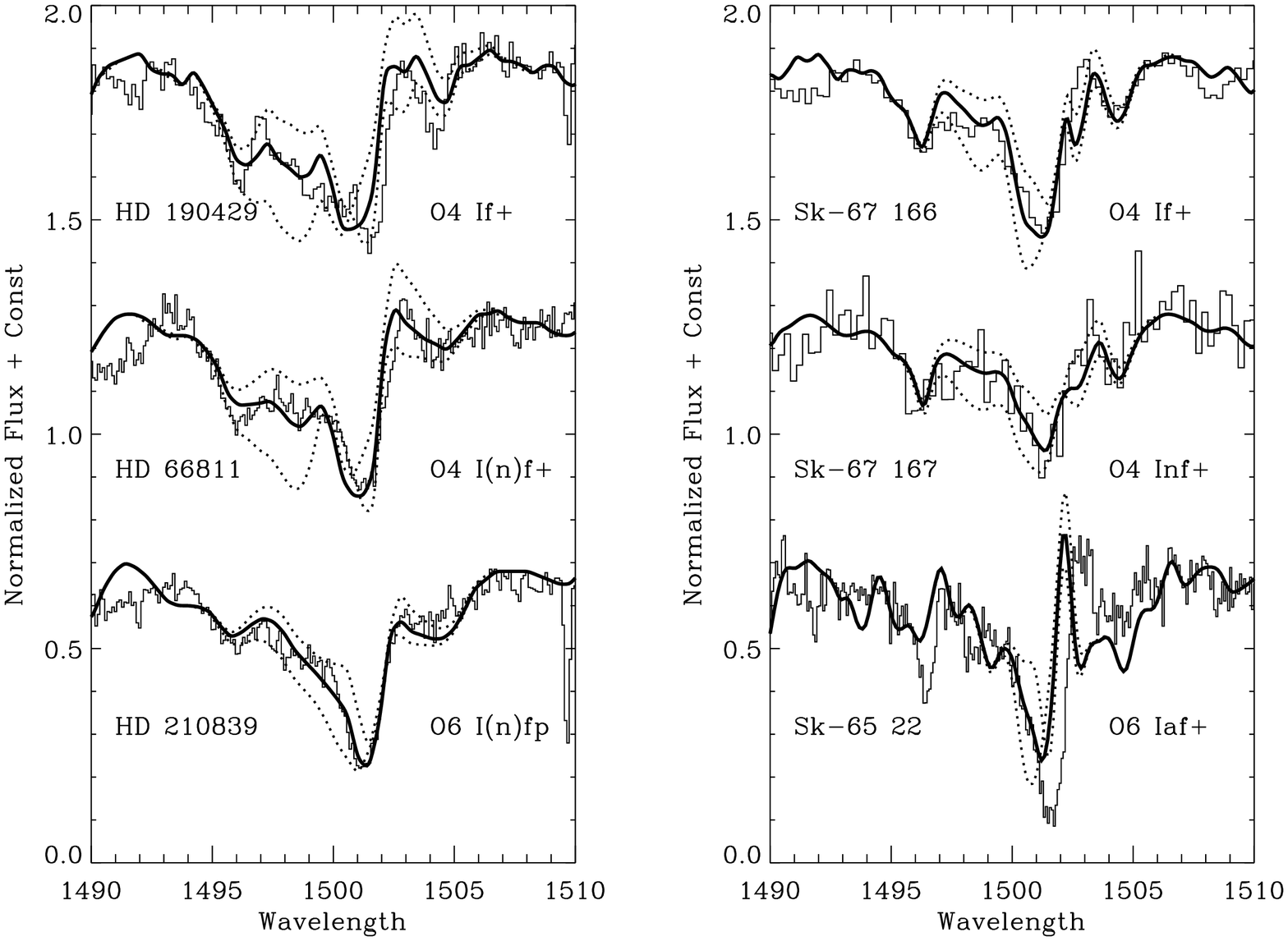,angle=90, width=4.9in}
\end{center}
\vspace{-0.2in}\noindent {\bf Figure 1:} Fits to the program stars.  The 
data are faint, solid curves, and the fits are heavy solid curves.  Models 
with $\tau_{rad}$ equal to twice and half of the best fit values are shown 
as dashed curves.  The values of $\dot{M}q(S^{+4})$ implied by the fits are 
between 0.04 and 0.18 of the expected values.

\end{document}